\documentclass[aps,prl,twocolumn,showpacs,10pt,superscriptaddress,preprintnumbers]{revtex4-1}
\usepackage{amsmath}
\usepackage{physics}
\usepackage{graphicx}
\usepackage{amssymb}
\usepackage[colorlinks,citecolor=red]{hyperref}

\usepackage{color}    
\usepackage{ulem}
\usepackage{multirow}

\newcommand{\beq}{\begin{equation}}
\newcommand{\eeq}{\end{equation}}
\newcommand{\bea}{\begin{eqnarray}}
\newcommand{\eea}{\end{eqnarray}}
\newcommand{\nn}{\nonumber}

\begin{document}

\preprint{
	{\vbox {
			\hbox{\bf LA-UR-21-26072}
			\hbox{\bf MSUHEP-21-014}
}}}
\vspace*{0.2cm}

\title{The anomalous $Zb\bar{b}$ couplings at the HERA and EIC}

\author{Bin Yan}
\email{binyan@lanl.gov}
\affiliation{Theoretical Division, Group T-2, MS B283, Los Alamos National Laboratory, P.O. Box 1663, Los Alamos, NM 87545, USA}

\author{Zhite Yu}
\email{yuzhite@msu.edu}
\affiliation{Department of Physics and Astronomy,
Michigan State University, East Lansing, MI 48824, USA}

\author{C.-P. Yuan}
\email{yuan@pa.msu.edu}
\affiliation{Department of Physics and Astronomy,
Michigan State University, East Lansing, MI 48824, USA}

\date{\today}

\begin{abstract}
To resolve the long-standing discrepancy between the precision measurement of bottom quark forward-backward asymmetry at LEP/SLC and the Standard Model prediction, we propose a novel method to probe the $Zb\bar{b}$ coupling by measuring the single-spin asymmetry $A_e^b$ of the polarized lepton cross section in neutral current DIS processes with a $b$-tagged jet at HERA and EIC. Depending on the tagging efficiency of the final state $b$-jet, the measurement of $A_e^b$ at HERA can already partially break the degeneracy found in the anomalous $Zb\bar{b}$ coupling, as implied by the LEP and SLC precision electroweak data. In the first year run of the EIC, the measurement of $A_e^b$ can already break the degeneracy, due to its much larger luminosity and higher electron beam polarization. With enough integrated luminosity collected at the EIC, it is possible to either verify or exclude the LEP data and resolve the $A_{\rm FB}^b$ puzzle. We also discuss the complementary roles between the proposed $A_e^b$ measurement at EIC and the measurement of $gg \to Zh$ cross section at the HL-LHC in constraining the anomalous $Zb\bar{b}$ coupling.

\end{abstract}

\maketitle

\noindent {\bf Introduction:~}
The bottom quark forward-backward asymmetry ($A_{\rm FB}^b$) at the $Z$-pole measured at the Large Electron-Positron collider (LEP) exhibits a long-standing discrepancy with the
Standard Model (SM) prediction by a significance 
around $2.1\sigma$~\cite{ParticleDataGroup:2020ssz}.
Though the observed discrepancy could be caused by statistical fluctuation or some subtle systematic errors in the experiment,  
it could be an evidence of new physics (NP) beyond the SM. 
Many global analyses have been carried out in the literature to include also the other experimental data sensitive to the $Zb\bar{b}$ coupling, such as the measurement of the branching fraction ($R_b$)  of  $Z\to b\bar{b}$ in the 
inclusive hadronic decay of $Z$ at LEP, and the left-right forward-backward asymmetry ($A_b$) of $b$ production at the Stanford Linear Collider (SLC) experiment with longitudinally polarized electron beam.
A class of popular NP models to interpret these precision data is 
to introduce an anomalous right-handed $Zb\bar{b}$ coupling, while keeping the left-handed $Zb\bar{b}$ coupling about the same as the SM~\cite{Choudhury:2001hs,Agashe:2006at,Gori:2015nqa,Liu:2017xmc}. Such a condition can be fulfilled for any theory with an underlying approximate custodial symmetry~\cite{Agashe:2006at}. 
However, it is well known that combining the $R_b$ and $A_{\rm FB}^b$, $A_b$ measurements at the $Z$-pole does not uniquely determine the $Zb\bar{b}$ coupling, but leads to 4 degenerate solutions~\cite{Choudhury:2001hs}. The LEP $A_{\rm FB}^b$ measurement off the $Z$-pole offers a way to break that degeneracy, but not completely, due to the low event rate, with still two possible solutions left~\cite{Choudhury:2001hs}. To break this degeneracy, more events off the $Z$-pole should be accumulated in a future lepton collider, such as 
CEPC, ILC, CLIC and FCC-ee~\cite{Gomez-Ceballos:2013zzn,Baer:2013cma,CEPCStudyGroup:2018ghi}. 

While still in wait for a next-generation lepton collider, we shall also seek for other opportunities that exist in the present or near future to break the above-mentioned degeneracy by, for example, probing exclusively the vector \textit{or} axial-vector component of the $Zb\bar{b}$ coupling, as well as to confirm or exclude the discrepancy between the $A_{\rm FB}^b$ measurement and the SM.
In a recent paper, we proposed to resolve  this puzzle through the precision measurement of $gg\to Zh$ scattering at the Large Hadron Collider (LHC)~\cite{Yan:2021veo}, 
which is sensitive to the axial-vector component of the $Zb\bar{b}$ coupling.
We showed that the high-luminosity LHC (HL-LHC) can not only break the degeneracy, but also verify or exclude 
the anomalous $Zb\bar{b}$ coupling observed at LEP through measuring the $Zh$ production rate, and this conclusion is not sensitive to possible NP  contribution induced by top quark or Higgs boson anomalous couplings in the loop~\cite{Yan:2021veo}.

In this paper we propose a novel method for directly measuring the vector, as opposed to the axial-vector, component of the $Zb\bar{b}$ coupling, to be carried out at lepton-hadron colliders,  
such as HERA and the upcoming Electron-Ion Collider (EIC).
The observable we propose
to probe the $Zb\bar{b}$ coupling is the single-spin asymmetry (SSA) $A_e^b$ of the polarized lepton-proton cross section in neutral current deeply-inelastic scattering (DIS) processes with one $b$-tagged jet in the final state~\cite{Chekanov:2009gm},
\beq
A_e^b= 
\frac{\sigma_{b,+}^{\rm tot} - \sigma_{b,-}^{\rm tot}}{\sigma_{b,+}^{\rm tot} + \sigma_{b,-}^{\rm tot}},
\label{eq:asy}
\eeq
where $ \sigma_{b,\pm}^{\rm tot} $ is the total inclusive $b$-tagged DIS cross section of 
a right-handed ($+1/2$) or left-handed ($-1/2$) lepton ({\it i.e.,} electron or positron) beam scattering off an 
unpolarized proton ($p$) beam.
In a parity-conserving theory, the SSA would be exactly zero. The photon-only diagrams will cancel in $A_e^b$, which accounts for the major portion of the cross section $\sigma_{b,\pm}^{\rm tot} $, such that we could gain direct probe to the diagrams with $Z$ boson exchange. To a very good approximation, the $Ze\bar{e}$ coupling can be taken to be fully axial-vector, for its vector component is almost zero. Hence, as to be shown below, only the vector component of the  $Zb\bar{b}$ coupling can contribute to the parity-violation  observable $A_e^b$ via the $\gamma$-$Z$ interference channel. 
When the momentum transfer scale $Q$ is much less than the $Z$ boson mass $m_Z$, the $Z$-$Z$ channel is suppressed by the $Z$ propagator at low $Q$. 
Therefore, the SSA is expected to give a sensitive probe to the  vector component of the  $Zb\bar{b}$  coupling. This information is complementary to that obtained by measuring the $gg\to Zh$ cross section at the LHC, which is  sensitive to the axial-vector component of the $Zb\bar{b}$ coupling~\cite{Yan:2021veo}.

We will demonstrate in the following that the asymmetry $A_e^b$ is indeed exclusively sensitive to the vector component of the $Zb\bar{b}$ coupling and the existing tensions and/or the degeneracy of the $Zb\bar{b}$ coupling from the electroweak precision data could be resolved by the $A_e^b$ measurements at the HERA and EIC.

\vspace{3mm}
\noindent {\bf DIS cross section:~}%
The polarized cross section for the scattering $e^\pm(k) + p(P)\to e^\pm(k^\prime) + X$ can be expressed in terms of structure functions~\cite{ParticleDataGroup:2020ssz} as 
\begin{align}
\frac{\dd\sigma_{\lambda_e}^\pm}{\sigma_0\dd x\dd y}=&F_1\left(\left(1-y\right)^2+1\right)+F_L\frac{1-y}{x}\mp F_3\lambda_e \left(y-\frac{y^2}{2}\right),
\end{align}
where $\sigma^{\pm}$ denotes the cross section for $e^{\pm}$, $\lambda_e=\pm 1$ denotes the helicity of incoming lepton, and $\sigma_0\equiv4\pi\alpha_{\rm em}^2/(xy^2S)$ with $S=(k+P)^2$ and $\alpha_{\rm em}$ the fine-structure constant. The standard DIS kinematic variables are defined as
\beq
Q^2=-q^2,\quad 
x=\frac{Q^2}{2P\cdot q},\quad 
y=\frac{P\cdot q}{P\cdot k},
\quad xyS=Q^2,
\eeq
where $q=k-k^\prime$ denotes the momentum transfer of the lepton. The structure functions $F_i\equiv F_i(x,y)$ with $i=1,2,3$ contain the contributions from photon-only channel $(F_i^\gamma)$, $Z$-only channel ($F_i^Z$) and the $\gamma$-$Z$ interference channel ($F_i^{\gamma Z}$), and can be written as
\begin{align}
F_i=F_i^\gamma - (g_V^e\pm\lambda_e g_A^e) \eta_{\gamma Z} F_i^{\gamma Z} +(g_V^e\pm\lambda_e g_A^e)^2 \eta_Z F_i^Z,
\end{align}
with the definition of $F_L\equiv F_2-xF_1$. The parameters $g_V^e=-1/2+2s_W^2$ and $g_A^e=-1/2$ are the vector and axial-vector couplings of electron to $Z$ boson, respectively. The factors $\eta_j$ denote the ratios of the couplings and propagators to the photon propagator and coupling, i.e.
\begin{align}
 &\eta_{\gamma Z}=\frac{Q^2}{Q^2+m_Z^2}\frac{1}{4c_W^2s_W^2},& \eta_Z=\eta_{\gamma Z}^2.
\end{align}
Here $s_W\equiv \sin\theta_W$ and $c_W\equiv\cos\theta_W$, with $\theta_W$ being the weak mixing angle, whose numerical value leads to small $g_V^e \simeq -0.038$, which justifies the approximation we made in the Introduction section. 

The structure functions with massless and massive quarks have been extensively discussed at the next-to-next-to-leading order (NNLO) accuracy in QCD~\cite{vanNeerven:1991nn,Zijlstra:1992qd,Zijlstra:1992kj,Larin:1996wd,Moch:1999eb,Guzzi:2011ew,Kawamura:2012cr}, and it is well known that the dependence of DIS cross section on heavy quark masses, $m_{c,b}$ can be significant when $Q\sim m_{c,b}$.  
Moreover, Wilson coefficients with massive quark lines must be calculated within the same factorization scheme as the one adopted to evaluate running of the strong coupling $\alpha_s$ and extract 
parton distributions functions (PDFs) from a global analysis of experimental data.
The Simplified-ACOT-$\chi$ (S-ACOT-$\chi$) scheme~\cite{Aivazis:1993kh,Aivazis:1993pi,Collins:1998rz,Kramer:2000hn,Tung:2001mv} has been employed successfully to calculate the heavy quark cross sections in the CTEQ PDF global analysis~\cite{Hou:2019efy,Dulat:2015mca}. 
In this study, we will discuss the impact of the HERA Run-II  data, with longitudinally polarized $e^\pm$ beam, which were not used in the determination of the CT14NNLO PDFs.~\cite{Dulat:2015mca}. 
Hence, in the following numerical calculation, we 
will take CT14NNLO PDFs  and  
calculate the structure functions $F_i$ with the  S-ACOT-$\chi$ scheme to the NNLO accuracy in QCD. 
In order to be consistent with the application of CT14NNLO PDFs, the pole mass $m_b=4.75~{\rm GeV}$ and $m_c=1.3~{\rm GeV}$ are used for calculating the massive structure functions.
The heavy quark structure functions contain contributions which are dependent on the $Z$-$q$-$\bar q$ coupling, with $q=c$ or $b$. 
At NNLO, heavy quark can also be produced in the final state of the subprocesses in which only light quark flavors are directly coupled to the $Z$ boson. 
This type of contribution to the total cross section is numerically small and has been consistently included in the light quark structure functions, as discussed in Ref.~\cite{Guzzi:2011ew}. 

\vspace{3mm}
\noindent{\bf Theoretical analysis:} Next, we discuss how the non-standard $Zb\bar{b}$ coupling could affect the value of SSA $A_e^b$, as defined in Eq.~\eqref{eq:asy}, measured at $e$-$p$ colliders. We parametrize the $Zb\bar{b}$ effective Lagrangian as,
\beq
\mathcal{L}_{\rm eff}=\frac{g_W}{2c_W}\bar{b}\gamma_\mu(\kappa_Vg_V^b-\kappa_Ag_A^b\gamma_5)b \, Z_\mu,
\eeq
where $g_W$ is the $SU(2)_L$ gauge coupling, and $g_V^b=-1/2+2/3s_W^2$ and $g_A^b=-1/2$ are the vector and axial-vector components of the $Zb\bar{b}$ coupling in the SM, respectively.
The coupling modifiers $\kappa_{V,A}$ are introduced to include possible NP effects in the $Zb\bar{b}$ interaction. A general analysis of NP effect on $A_e^b$, including all possible higher dimensional operators, will be presented elsewhere.

We now give a detailed analysis on how $\kappa_V$ and $\kappa_A$ can appear in the SSA $A_e^b$. 
First of all, the numerator of  Eq.~\eqref{eq:asy} takes a form of helicity difference. Taking helicity difference of the incoming lepton is equivalent to inserting one $\gamma_5$ matrix.
Hence, a non-zero $A_e^b$ can only occur with the presence of some 
parity-violating structures 
(involving $Z$ boson couplings)
in the diagrams; specifically, they can be
\begin{itemize}
\item in the $\gamma Z$ interference diagrams,
	$$g_A^e \cdot \kappa_V, \mbox{ or } g_V^e \cdot \kappa_A;$$
\item and in the $Z$-only diagrams,
	$$\left( (g_V^e)^2 + (g_A^e)^2 \right) \cdot \kappa_V \kappa_A, \mbox{ or } 
		g_V^e g_A^e \cdot \left( \kappa_V^2 + \kappa_A^2 \right).$$
\end{itemize}
where $\kappa_{V,A}$ should be interpreted as being multiplied by $g_{V,A}^b$, respectively. Since $g_V^e \ll g_A^e$ and $Z$-only diagrams are suppressed by one more power of $Z$ propagator, the most dominant term is $g_A^e \cdot \kappa_V$ in the $\gamma Z$ interference diagrams. Furthermore, the $g_V^e \cdot \kappa_A$ term in $\gamma Z$ channel and $\kappa_V \kappa_A$ term in $ZZ$ channel are associated with $F_3$ structure function, which involves the convolution of the difference between quark and antiquark PDFs $f_q - f_{\bar{q}}$. Both at leading-order (LO) and next-to-leading order (NLO), the quark parton out of the proton is identical to the quark directly interacting with the vector boson in the Feynman diagrams, so $q = b$, and $f_b - f_{\bar{b}} = 0$ exactly. Hence, those terms linear with $\kappa_A$ vanish up to NLO. At NNLO, $q$ can be other quark flavor, and $f_b - f_{\bar{b}}$ is not zero due to DGLAP evolution effect~\cite{Moch:2004pa,Catani:2004nc}, and thus the contribution from the $\kappa_A$-linear terms starts to be non-zero, though small. Therefore, we conclude that 
\textit{the numerator in Eq.~\eqref{eq:asy} dominantly depends linearly on $\kappa_V$ through $\gamma Z$ interference diagrams.} 
The dependence on $\kappa_A$ is suppressed by small $g_V^e$, or $Z$ propagator, or  higher-order effects. The denominator in Eq.~\eqref{eq:asy} is dominated by $\gamma$-only diagrams, and does not affect this conclusion.  In the following numerical analysis, we have included the full NNLO QCD corrections, with both the $\kappa_V$ and $\kappa_A$ dependence included, and verified that the contribution from $\kappa_A$ to $A_e^b$ is negligible.
 
We note that although we are interested in comparing to the $b$-tagged cross section only, both light and heavy quarks can contribute to the measured value of $A_e^b$ in practice. This arises from the imperfect jet tagging efficiency, which is dependent on the jet transverse momentum and rapidity observed in the final state.
It has been shown at the LHC~\cite{Chatrchyan:2012jua} that when the $b$-jet tagging efficiency $(\epsilon_b)$ is within [0.5,~0.7], the misidentification probability of a light quark as a $b$-jet $(\epsilon_q^b)$ can be $[10^{-3},~10^{-2}]$, and the misidentification probability of a $c$ quark as a $b$-jet $(\epsilon_c^b)$ is in [0.03,~0.2]. Importantly, these probabilities are not very sensitive to the kinematics of the jets~\cite{Chatrchyan:2012jua}. Therefore,
although the detectors and jets' kinematic distributions are different between the LHC and HERA or EIC, the tagging efficiencies at the LHC can serve as a good reference that can be used to estimate the expected sensitivity of the $Zb\bar{b}$ coupling from the $A_e^b$ measurement at HERA or EIC. In this study, we  hence adopt two sets of benchmark tagging efficiencies as
\begin{align}
(i)~ & \epsilon_q^b=0.001,&&\epsilon_c^b=0.03,&&\epsilon_b=0.7;\nn\\
(ii)~ & \epsilon_q^b=0.01,&&\epsilon_c^b=0.2,&&\epsilon_b=0.5. 
\label{eq:eff}
\end{align}
Here, the scenario $(i)$  represents a good $b$-tagging efficiency, and $(ii)$ a worse one. 

Taking into account all possible quark flavor contributions,  the $b$-tagged inclusive cross section can be approximated by
\beq
\sigma_b^{\rm tot}(\lambda_e)=\sum_{q=u,d,s}\sigma_q(\lambda_e)\epsilon_q^b+\sigma_c(\lambda_e)\epsilon_c^b+\sigma_b(\lambda_e,\kappa_V,\kappa_A)\epsilon_b.
\eeq
Below, we present the  numerical result of our calculation, using CT14NNLO PDFs~\cite{Dulat:2015mca} for HERA $(E_{\rm cm}=318~{\rm GeV})$ and EIC $(E_{\rm cm}=141~{\rm GeV})$, up to the NNLO accuracy. Both the renormalization and factorization scales are fixed at $\mu=Q$.  

\setlength{\tabcolsep}{8pt}
\begin{table}
	\begin{center}
		\begin{tabular}{r|c|c}
			\hline
			H1& $R$ & $L$\\
			\hline
			$e^-p$ 
			& $47.3\,{\rm pb}^{-1}$, $0.36$  & $104.4\,{\rm pb}^{-1}$, $-0.258$  \\
			\hline
			$e^+p$
			& $101.3\,{\rm pb}^{-1}$, $0.325$ & $80.7\,{\rm pb}^{-1}$, $-0.37$  \\
			\hline
			\hline
			ZEUS& $R$ & $L$\\
			\hline
			$e^-p$
			& $71.2\,{\rm pb}^{-1}$, $0.29$ & $98.7\,{\rm pb}^{-1}$, $-0.27$ \\
			\hline
			$e^+p$  
			& $78.8\,{\rm pb}^{-1}$, $0.32$ & $56.7\,{\rm pb}^{-1}$, $-0.36$ \\
			\hline
		\end{tabular} 
		\caption{The integrated luminosity and lepton beam's longitudinal polarization for each data set of H1~\cite{Aaron:2012qi} and ZEUS~\cite{Chekanov:2009gm,Abramowicz:2012bx}. $R$ ($L$) denotes right-handed (left-handed) lepton data set.}
		\label{tab:lumi}
	\end{center}
\end{table}

\vspace{3mm}
\noindent{\bf Sensitivity at the HERA:} 
The HERA experiment used both electron and positron beams, with different degrees of polarization and luminosities. The polarization and luminosity also differ between right-handed and left-handed data sets, as shown in Table~\ref{tab:lumi}. The SSA in Eq.~\eqref{eq:asy} is related to experimental measurement by~\cite{Chekanov:2009gm}
\beq
A_e^b = 
\frac{\sigma_b^{\rm tot}(P_e)-\sigma_b^{\rm tot}(-P_e^\prime)}{P_e^\prime \,\sigma_b^{\rm tot}(P_e) + P_e\,\sigma_b^{\rm tot}(-P_e^\prime)},
\label{eq:asyHera}
\eeq
where $\sigma_b^{\rm tot}(P_e)$ denotes the total $b$-tagged cross section measured in the experiment  for the lepton beam with polarization $P_e$, and we take the convention $P_e, P_e' > 0$.
In this study, we focus on the following kinematic regions: $x\in [0.002,0.65]$ and $Q^2\in [120,50000]~{\rm GeV}^2$ for H1~\cite{Aaron:2012qi}, and $x\in [0.004,0.75]$ and  $Q^2\in [185,50000]~{\rm GeV}^2$ for ZEUS~\cite{Chekanov:2009gm,Abramowicz:2012bx} experiments, respectively. 

The systematic uncertainties are assumed to cancel in SSA and can be ignored~\cite{Chekanov:2009gm}. 
With the tagging efficiencies considered in Eq.~\eqref{eq:eff}, we 
calculate the relative error  $\delta A_e^b / A_e^b$ and find it to be  
around $60\%\sim80\%$ for each HERA experiment listed in Table~\ref{tab:lumi}.
To estimate the impact from the H1 and ZEUS data on constraining 
the anomalous $Zb\bar{b}$ coupling, 
we conduct a combined $\chi^2$ analysis with 
all the HERA II data sets, cf. Table~\ref{tab:lumi},  included as 
\beq
\chi^2=\sum_i\left[\frac{(A_e^b)_i^{\rm th}-(A_e^b)_i^{\rm exp}}{\delta A_e^b}\right]^2,
\eeq
where $(A_e^b)_i^{\rm th}$ and $(A_e^b)_i^{\rm exp}$ are, respectively, the theoretical predictions at the NNLO accuracy and the  experimental values for the $i$-th asymmetry data. In this study, for simplicity, we assume that the experimental values agree with the SM prediction, {\it i.e.,} $(A_e^b)_i^{\rm exp}=(A_e^{b0})_i$.

 \begin{figure}
	\includegraphics[scale=0.26]{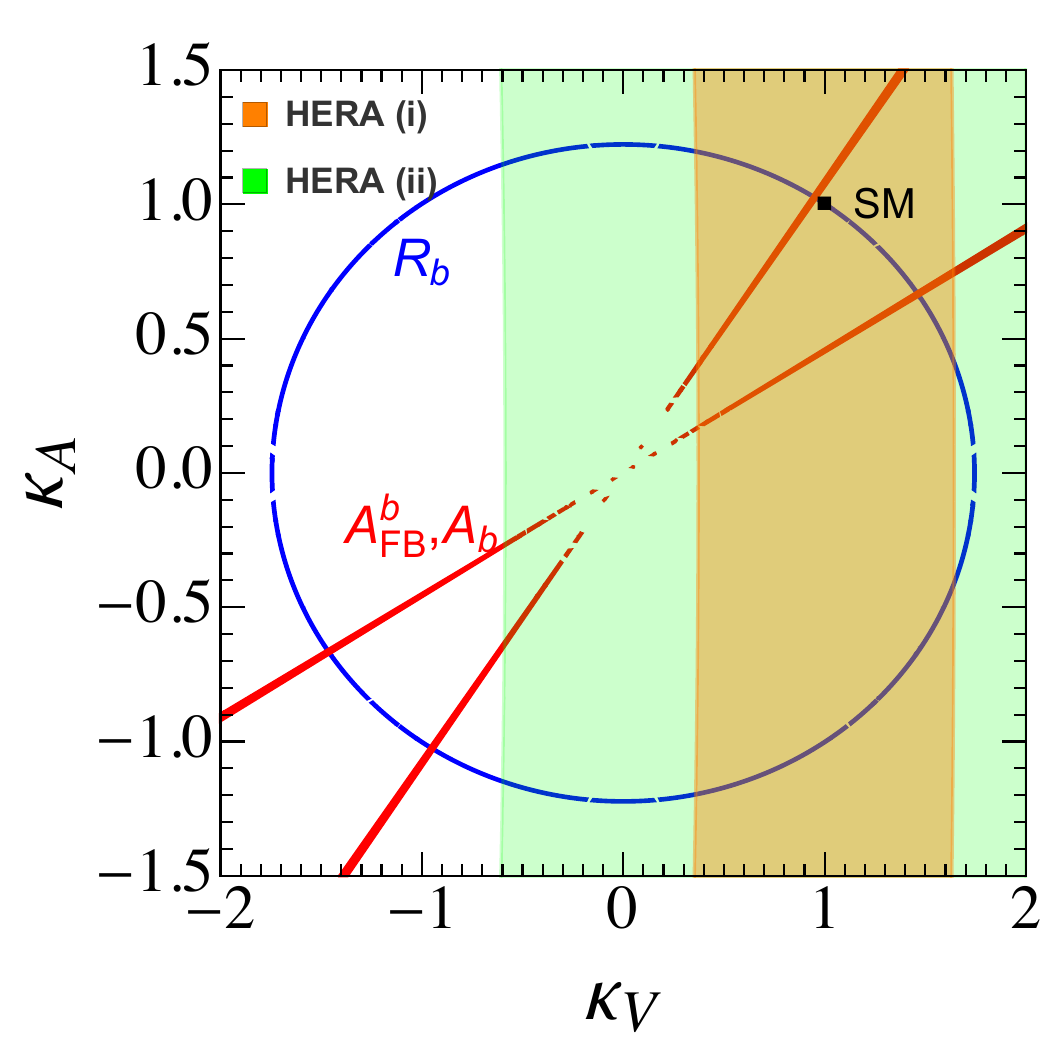}
	\caption{The expected limits on the anomalous $Zb\bar{b}$ couplings $\kappa_V$ and $\kappa_A$ from the $A_e^b$ measurements at HERA, at  68\% C.L.. The blue and red regions come from the $R_b$ and $(A_{\rm FB}^b, A_b)$ measurements at the LEP and SLC, respectively. The orange and green bands come from the measurements of the asymmetry $A_e^b$ at HERA,  with the tagging efficiency scenarios $(i)$ and $(ii)$ in  Eq.~\eqref{eq:eff}, respectively. }
\label{Fig:HERA}
\end{figure}

In Fig.~\ref{Fig:HERA}, we compare the precision of  determining the $Zb\bar{b}$ coupling through the $A_e^b$ measurements at HERA to the precision electroweak data at LEP and SLC. The blue and red shaded regions denote the constraints, at 68\% confidence level (C.L.),  from the $R_b$ and $(A_{\rm FB}^b,A_b)$ measurements at the $Z$-pole, respectively. The orange and green bands correspond to the measurements of SSA $A_e^b$ at HERA with the tagging efficiencies $(i)$ and $(ii)$ in Eq.~\eqref{eq:eff}, respectively. 
It is evident from the orange and green bands in  Fig.~\ref{Fig:HERA} that the measurement of $A_e^b$ is sensitive to the vector component of the $Zb\bar{b}$ coupling $\kappa_V$, but not the axial-vector component $\kappa_A$, as argued before.
This is opposite to 
the impact of measuring the production rate of $gg\to Zh$ at  LHC~\cite{Yan:2021veo}, which 
is sensitive to  $\kappa_A$, but not  $\kappa_V$. 
As a result, the HERA (and EIC, to be discussed below) are complementary to the LHC and lepton colliders in the measurement of the $Zb\bar{b}$ coupling.

From Fig.~\ref{Fig:HERA}, it is clear that two of the degenerate solutions with $\kappa_{V,A}<0$ could be excluded after combining the HERA data with the $Z$-pole measurements, and this conclusion still holds even when we use a worse tagging efficiency (green band). Therefore, the HERA data could be used to crosscheck the off-$Z$-pole $A_{\rm FB}^b$ measurement at the LEP, which has been used to exclude the part of the parameter space with $\kappa_{V,A}<0$~\cite{Choudhury:2001hs}. However, it remains difficult to resolve the apparent degeneracy in the parameter space with $\kappa_{V,A}>0$ due to the limited statistics for both the  off-$Z$-pole data at LEP and the $A_e^b$ measurement at HERA. 
In order to break the 
remaining degeneracy with $\kappa_{V,A}>0$ 
by the $A_e^b$ measurement, it requires a much better  jet-tagging efficiency at HERA, e.g.,  $\epsilon_q^b=0.0005,~\epsilon_q^c=0.01,$ and ~$\epsilon_q^b=0.95$.

\begin{figure}
	\includegraphics[scale=0.24]{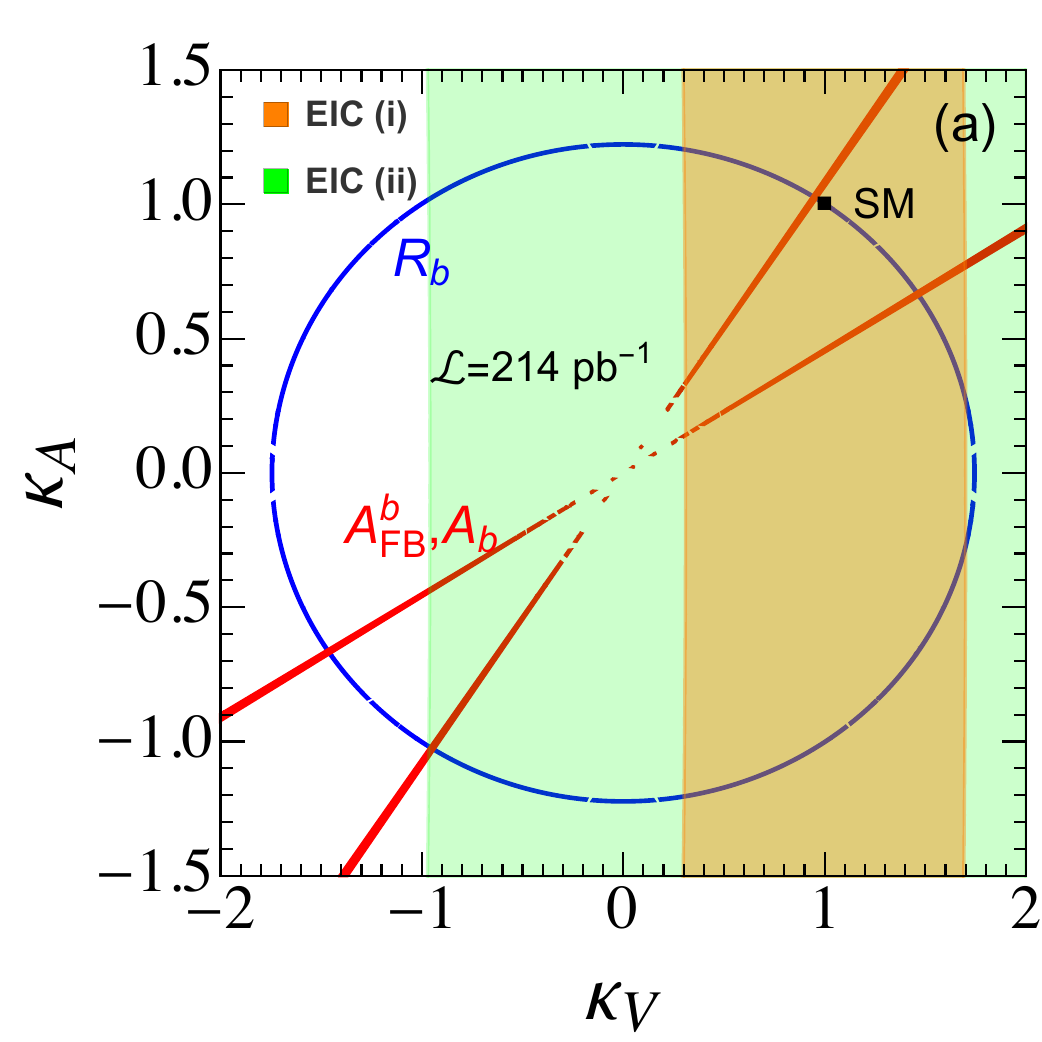}
	\includegraphics[scale=0.24]{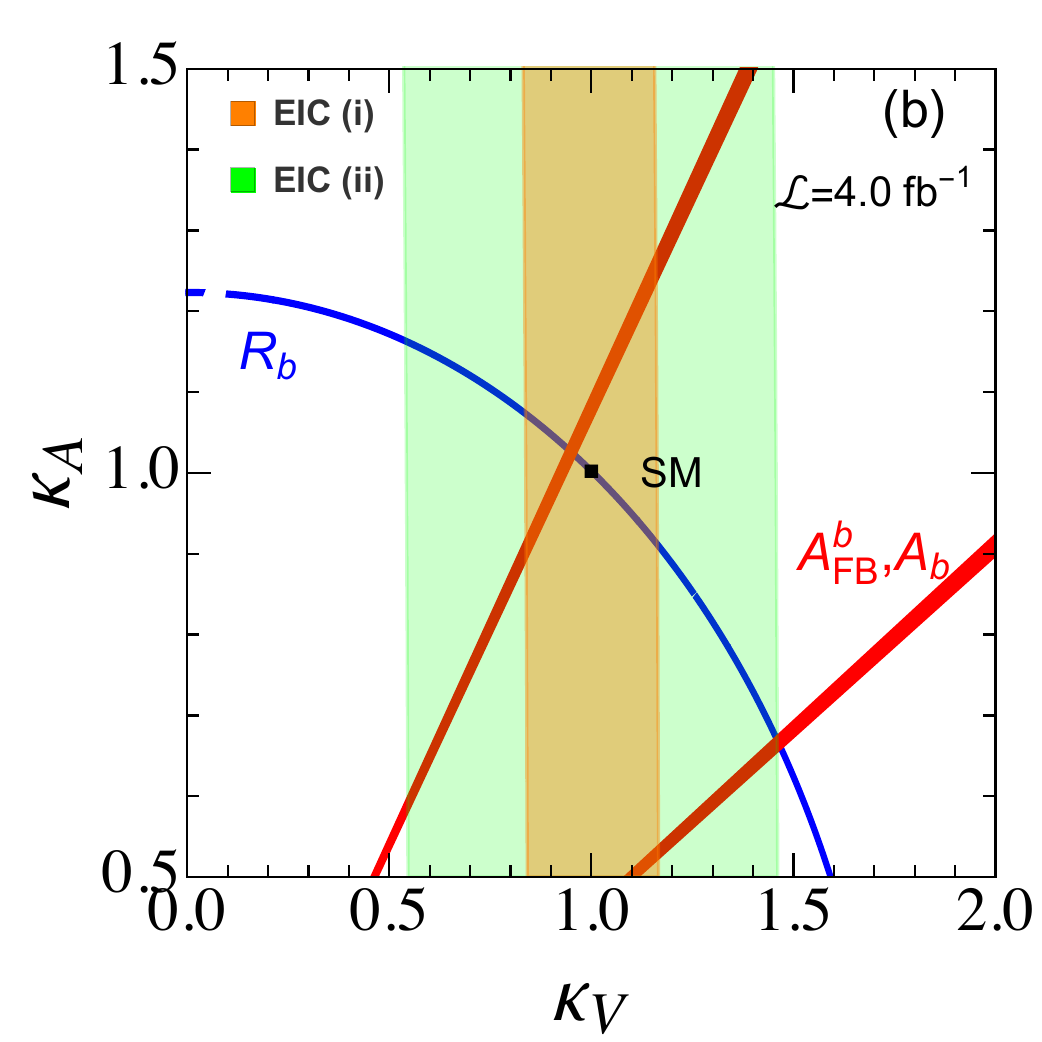}
	\includegraphics[scale=0.24]{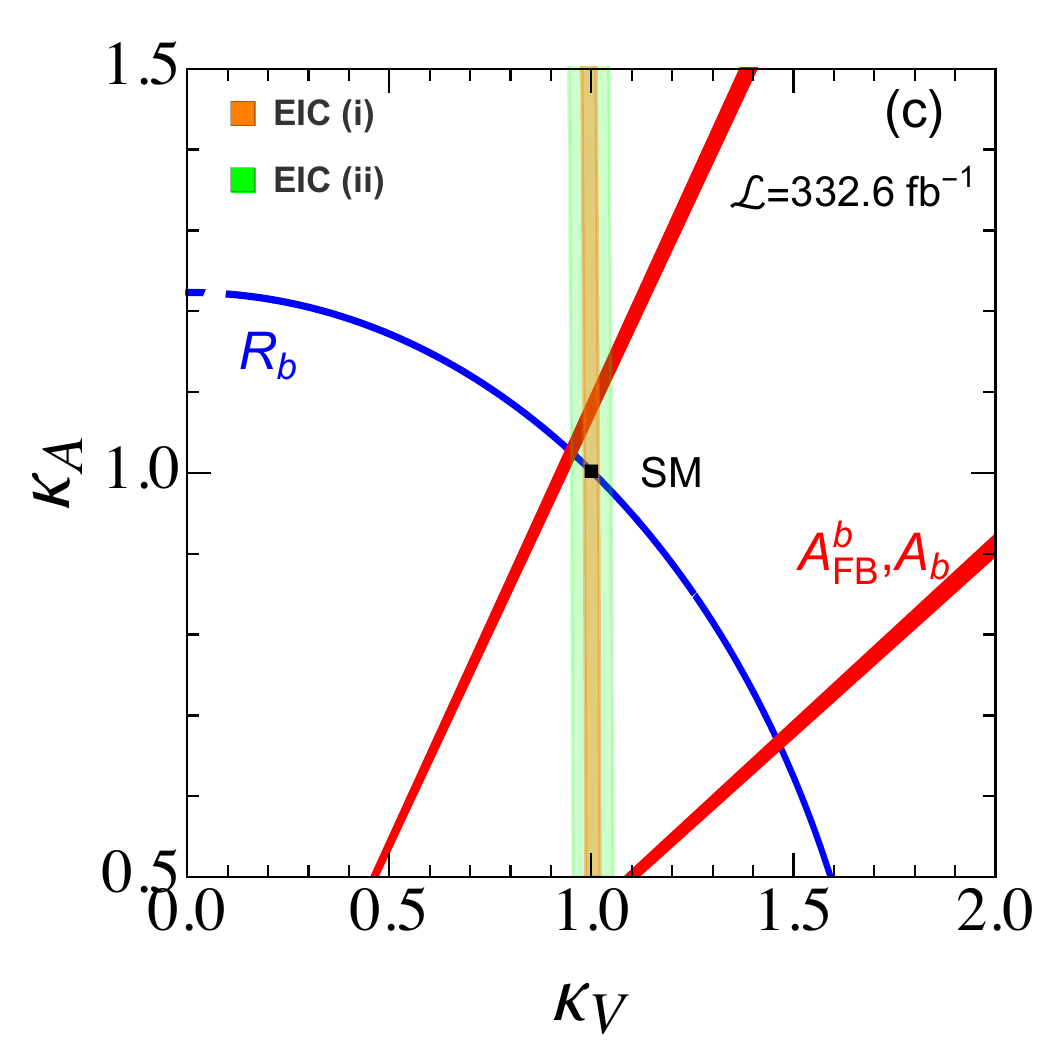}
	\caption{
Similar to  Fig.~\ref{Fig:HERA}, but at the EIC, with $E_{\rm cm}=141~{\rm GeV}$ and $P_e = 70\%$.}
	\label{Fig:EIC}
\end{figure}

\vspace{3mm}
\noindent{\bf Sensitivity at the EIC:} 
The upcoming EIC has a lower center-of-mass energy than HERA, but with a vastly higher luminosity and beam polarization, reaching $10-100~{\rm fb}^{-1}$ per year, with polarization as high as $70\%$~\cite{Accardi:2012qut}. We only consider electron beams at EIC, and the degrees of polarization and integrated luminosities for right- and left-handed electron beams are taken to be the same. Then Eq.~\eqref{eq:asy} is translated to experimental measurement via
\beq
A_e^b = \frac{1}{P_e}
\frac{\sigma_b^{\rm tot}(P_e)-\sigma_b^{\rm tot}(-P_e)}{\sigma_b^{\rm tot}(P_e) + \sigma_b^{\rm tot}(-P_e)},
\eeq
with $\sigma_b^{\rm tot}(P_e)$ having the same meaning as in Eq.~\eqref{eq:asyHera}, and $P_e = 70\%$ at EIC. The statistical uncertainty of $A_e^b$ is
\begin{align}
\delta A_e^b = 
\sqrt{\frac{P_e^{-2} - \left(A_e^b\right)^2}{\mathcal{L}\cdot\left[ \sigma_{b0}^{\rm tot}(P_e) + \sigma_{b0}^{\rm tot}(-P_e) \right]}}
\simeq 
\frac{1 / P_e}{\sqrt{2 \, \mathcal{L} \, \sigma_{b0}^{\rm tot}(0)}},
\end{align}
where the subscript `0' 
indicates the SM prediction, 
{\it i.e.,} with $\kappa_V = \kappa_A = 1$, and the second equality takes the approximation of $A_e^b \ll 1$, with $\sigma_{b0}^{\rm tot}(0)$ denoting the unpolarized $b$-tagged cross section. So with a higher luminosity and beam polarization, EIC shall give a much stronger constraint on the $Zb\bar{b}$ coupling than HERA.

To improve jet reconstruction in the hadronic final state and to 
enhance the contribution of $\gamma Z$ interference channel, which
dominates the SSA $A_e^b$, 
we only consider data with $Q>10~{\rm GeV}$. Thus, we shall focus on the EIC  kinematic region:   $x\in[0.005,0.8]$ and $Q^2\in [10^2,10^4]~{\rm GeV}^2$~\cite{AbdulKhalek:2021gbh}. 
With the canonical integrated luminosity,  $\mathcal{L} = 10~{\rm fb}^{-1}$,  and the tagging efficiencies ($i$ and $ii$) in Eq.~\eqref{eq:eff}, the SM predictions yield $(i) A_e^b = -0.023$, $\delta A_e^b/ A_e^b = 7.7\%$; $(ii) A_e^b = -0.014$, $\delta A_e^b/ A_e^b = 6.5\%$.

In Fig.~\ref{Fig:EIC}, we show the expected constraining power of EIC on the anomalous $Zb\bar{b}$ coupling $\kappa_V$ and $\kappa_A$ with different luminosities at 68\% C.L.. 
We find that the minimal luminosities needed to exclude the degeneracy in the parameter space with $\kappa_{V,A}<0$,  under the two choices ($i$ and $ii$) of the tagging efficiencies listed in Eq.~\eqref{eq:eff} are, respectively, (see Fig.~\ref{Fig:EIC}(a))
\beq
(i): \mathcal{L}>27~{\rm pb}^{-1};\quad\quad (ii): \mathcal{L}>214~{\rm pb}^{-1}.
\eeq
The minimal luminosities needed to resolve the apparent degeneracy in the parameter space with $\kappa_{V,A}>0$, {\it i.e.,}  ($\kappa_V,\kappa_A$)=(1.46,0.67),  are, respectively,  (see Fig.~\ref{Fig:EIC}(b))
\beq
(i): \mathcal{L}>0.5~{\rm fb}^{-1};\quad (ii): \mathcal{L}>4.0~{\rm fb}^{-1}.
\eeq
The minimal luminosities to exclude the LEP $A_{\rm FB}^b$ measurements, 
{\it i.e.,} to exclude the solution that is close to the SM,  are, respectively, (see Fig.~\ref{Fig:EIC}(c))
\beq
(i): \mathcal{L}>42.0~{\rm fb}^{-1};\quad (ii): \mathcal{L}>332.6~{\rm fb}^{-1}.
\eeq
Hence, with the ``canonical'' integrated luminosity, about 10 fb$^{-1}$, the measurement of $A_e^b$ at EIC can already break the degeneracy in the first year of running. 
In comparison with the potential of the HL-LHC to break the degeneracy, EIC is surely a better machine for this task, unless a large deviation from the SM coupling is found experimentally. 
To exclude the LEP $A_{\rm FB}^b$ measurements would probably require the maximal luminosity scenario of the EIC, over a few years of running~\cite{Accardi:2012qut}.

Finally, we remark that though we focus on final states with b-tagged jet in this study, the same method can also be applied to final states with $b$-hadrons. Furthermore, a similar analysis can also be applied to the Large Hadron Electron Collider (LHeC)~\cite{LHeCStudyGroup:2012zhm} at CERN to constrain the $Zb\bar{b}$ anomalous couplings.

\vspace{3mm}
\noindent {\bf Conclusions:~}%
In this Letter, we propose to probe the $Zb\bar{b}$ coupling by measuring the 
single-spin asymmetry $A_e^b$ of the polarized lepton cross section in neutral current DIS processes with a $b$-tagged jet at HERA and EIC. 
We show that $A_e^b$ is exclusively sensitive to the vector component of the $Zb\bar{b}$ coupling, and plays a complementary role to the measurement of the total cross section of $gg \to Zh$ at the HL-LHC, which is sensitive to the axial-vector component of the $Zb\bar{b}$ coupling.
Depending on the tagging efficiency of the final state $b$-jet,  
the measurement of $A_e^b$ at HERA can already partially break the degeneracy 
found in the $Zb\bar{b}$ coupling, as implied by the LEP and SLC precision electroweak data. It takes EIC to clarify the long-standing discrepancy between the LEP $A_{\rm FB}^b$ data and the SM prediction, because of its much higher luminosity and electron beam polarization.
With enough integrated luminosity collected at the EIC, it is possible to either verify or exclude the LEP data and resolve the $A_{\rm FB}^b$
puzzle.  

\vspace{3mm}
\noindent{\bf Acknowledgments.}
The authors thank Yu-Xiang Zhao for helpful discussion. This work is partially supported by the U.S. Department of Energy, Office of Science, Office of Nuclear Physics, under Contract DE-AC52-06NA25396 through the LANL/LDRD Program, as well as the U.S.~National Science Foundation
under Grant No.~PHY-2013791. C.-P.~Yuan is also grateful for the support from the Wu-Ki Tung endowed chair in particle physics.

\bibliographystyle{apsrev}
\bibliography{reference}

\end{document}